\documentstyle[12pt,aasms4]{article}
\received{July 15, 1999}
\accepted{October 25, 1999}
\slugcomment{Accepted for Publication in The Astronomical Journal}
\lefthead{Eskridge et al.}
\righthead{Infra-Red Bars}

\begin{document}

\title{The Frequency of Barred Spiral Galaxies in the Near-IR\footnotemark}

\author{Paul B.~Eskridge\altaffilmark{2}, Jay A.~Frogel\altaffilmark{2,3}, 
Richard W.~Pogge\altaffilmark{2}, Alice C.~Quillen\altaffilmark{3,4}, 
Roger L.~Davies\altaffilmark{5}, D.L.~DePoy\altaffilmark{2,3}, Mark 
L.~Houdashelt\altaffilmark{6}, Leslie E.~Kuchinski\altaffilmark{3,7}, Solange 
V.~Ram\'{\i}rez\altaffilmark{2,3}, K.~Sellgren\altaffilmark{2,3}, Donald 
M.~Terndrup\altaffilmark{2,3} \& Glenn P.~Tiede\altaffilmark{3,8}}

\footnotetext{Based partially on observations obtained at the Cerro Tololo
Interamerican Observatory, operated by the Association of Universities for
Research in Astronomy, Inc.~(AURA) under cooperative agreement with the
National Science Foundation}
\altaffiltext{2}{Department of  Astronomy, The Ohio State University, Columbus, 
OH 43210}
\altaffiltext{3}{Visiting Astronomer, Cerro Tololo Interamerican Observatory}
\altaffiltext{4}{Steward Observatory, University of Arizona, Tucson AZ 85721}
\altaffiltext{5}{Department of Physics, South Road, University of Durham, 
Durham DH1 3LE, England}
\altaffiltext{6}{Department of Physics and Astronomy, Johns Hopkins University, 
Baltimore, MD 21218}
\altaffiltext{7}{NASA/IPAC, California Institute of Technology, MS 100-22, 
Pasadena, CA 91125}
\altaffiltext{8}{NOAO, Tucson, AZ  85719}

\authoremail{eskridge@astronomy.ohio-state.edu, frogel@astronomy.ohio-state.edu}

\begin{abstract}
We have determined the fraction of barred galaxies in the $H$-band for a 
statistically well-defined sample of 186 spirals drawn from the Ohio State
University Bright Spiral Galaxy survey.  We find 56\% of our sample to be 
strongly barred at $H$, while another 16\% is weakly barred.  Only 27\% of our 
sample is unbarred in the near-infrared.  The RC3 and the Carnegie Atlas of 
Galaxies both classify only about 30\% of our sample as strongly barred.  Thus 
{\it strong} bars are nearly twice as prevalent in the near-infrared as in the 
optical.  The frequency of genuine optically hidden bars is significant, but 
lower than many claims in the literature:  40\% of the galaxies in our sample 
that are classified as unbarred in the RC3 show evidence for a bar in the 
$H$-band, while for the Carnegie Atlas this fraction is 66\%.  Our data reveal 
no significant trend in bar fraction as a function of morphology in either the 
optical or $H$-band.  Optical surveys of high redshift galaxies may be strongly 
biased against finding bars, as bars are increasingly difficult to detect at 
bluer rest wavelengths.
\end{abstract}


\section{Introduction}

Roughly 30\% of spiral galaxies are strongly barred in the optical 
(\markcite{old}de Vaucouleurs 1963), while another 25\% are weakly barred 
(e.g., \markcite{sw}Sellwood \& Wilkinson 1993).  Evidence that bars in spiral 
galaxies are more obvious in the near-infrared (IR) than the visible goes back 
to \markcite{hs}Hackwell \& Schweizer (1983), who noted that the weak optical 
bar in NGC 1566 was a strong feature in the $H$-band, with ``azimuthal 
brightness variations of up to $\sim$25\%.''  Their main conclusion was that 
these brightness variations reflected true mass variations.  The case for this 
interpretation is reviewed by \markcite{fqp}Frogel, Quillen \& Pogge (1996).  
Briefly, the main reasons are that the $H$-band light is sampling the dominant, 
old stellar population and is relatively unaffected by the presence of bright 
young blue stars or by dust, both of which strongly influence optical images.  
Since \markcite{hs}Hackwell \& Schweizer's initial discovery, several other 
examples of near-IR bars that are less obvious in the optical have been 
reported (e.g., \markcite{hfd}Thronson et al.~1989 -- NGC 1068; 
\markcite{bw}Block \& Wainscoat 1991 -- NGC 309; \markcite{bea}Block et 
al.~1994 -- NGC 4736; \markcite{mnr}Mulchaey \& Regan 1997; 
\markcite{snj}Seigar \& James 1998; \markcite{ksp}Knapen, Shlosman \& Peletier 
1999).  While useful in pointing out specific cases of optically hidden bars, 
none of these near-IR studies are based on large, statistically well-defined 
samples of galaxies, and so cannot address the following fundamental question:  
What fraction of spiral galaxies have bars?  

A determination of the true frequency of bars and other nonaxisymmetric 
structures is important for a number of reasons.  Bars can initiate spiral 
density waves (e.g., \markcite{ato}Toomre 1969; \markcite{k8a}Kormendy 1982a) 
and stellar rings (\markcite{b86}Buta 1986).  Stellar bars may also provide a 
means to channel gas into nuclear regions of galaxies by inducing gravitational 
torques on gas trailing the bar, thus robbing the gas of angular momentum and 
driving it inward to fuel nuclear starbursts or AGN (e.g., 
\markcite{k8b}Kormendy 1982b; \markcite{sfb}Shlosman, Frank \& Begelman 1989; 
\markcite{pfn}Pfenniger \& Norman 1990; \markcite{qea}Quillen et al.~1995; 
\markcite{pst}Piner, Stone \& Teuben 1995).  Finally, bars may be important for 
driving the secular evolution of bulges, both by triggering nuclear starbursts 
and by kinematic heating of the inner disk (\markcite{pfn}Pfenniger \& Norman 
1990; \markcite{cea}Combes et al.~1990).  

A number of studies indicate that strongly barred galaxies must have maximal 
(or nearly maximal) disks.  \markcite{dbs}DeBattista \& Sellwood (1998) argue 
that bars in submaximal disks should have short lifetimes, as they will give up 
their angular momentum to their halos.  Also, bars are often associated with 
rings (e.g., \markcite{src}Buta 1995).  Such bar/ring structures appear to 
require a maximal disk for stability (\markcite{qnf}Quillen \& Frogel 1997).  
Hydrodynamic simulations often find that the gas dynamics in barred systems 
require massive bars (e.g., \markcite{lla}Lindblad, Lindblad \& Athanassoula 
1996; \markcite{wea}Weiner et al.~1996).  The situation is strikingly different 
for galaxies that are optically unbarred.  \markcite{cnr}Courteau \& Rix (1999) 
find that the Tully-Fisher relationship shows no dependence on disk scale 
length for a sample of optically unbarred galaxies, and conclude from this that 
disks are massless.  \markcite{b93}Bottema (1993) studied the velocity 
dispersion profiles of a sample of optically unbarred galaxies, and concluded 
that these galaxies are all submaximal.  To make matters more puzzling, there 
does not appear to be any difference in the average rotation curves of 
optically barred and unbarred galaxies (\markcite{b96}Bosma 1996).  Thus a 
determination of the true frequency of bars (hereafter `bar fraction') in 
spiral galaxies is essential for understanding the so-called ``dark matter 
problem,'' and, more generally, the structure of spiral galaxy halos. 

In this paper we present the first statistical study of the frequency of bars 
in the near-IR based on the Ohio State Bright Spiral Galaxy Survey.  In \S 2 we 
briefly describe the survey.  We examine the near-IR bar fraction of our sample
in \S 3, and we compare this to the results for the same sample from standard 
optical catalogs.  Our sample is large enough that we can study the effect of 
morphology on bar statistics in \S 4.  In \S 5 we discuss some implications of 
our results, consider the possibility of various biases in our analysis, and 
indicate promising areas for future research.

\section{The OSU Bright Spiral Galaxy Survey}

The Ohio State Bright Spiral Galaxy Survey comprises deep, photometrically 
calibrated $BVRJHK$ images of a magnitude-limited sample of 205 spirals with $T 
\geq 0$, $B \leq 12$ and $D \leq 6'$ in \markcite{rc3}de Vaucouleurs et 
al.~(1991, hereafter RC3).  A full description of the survey, including the 
selection criteria and observational strategy, will be presented in 
\markcite{bap}Eskridge et al.~(1999).  Because this is a large, clearly defined 
sample of well-resolved galaxies, it is especially useful for addressing the 
statistical properties of spiral galaxies.  

Although the data from the Ohio State Survey are not yet fully photometrically
calibrated, they are well suited for performing the first statistical analysis 
of the presence of bars in near-IR images of spiral galaxies.  We base our 
analysis on our $H$-band images, as they generally have higher signal to noise 
than do our $K$-band images and are nearly as unaffected by dust extinction.  
For this paper, we consider only galaxies that do not display gross 
peculiarities, and do not have close companions of similar (or greater) 
luminosity.  This gives us a total sample of 186 galaxies.

\section{Bar Frequency in the Near-IR and Optical}

As our sample is selected from the \markcite{rc3}RC3, we can compare our
$H$-band bar classifications with the \markcite{rc3}RC3 optical bar
classifications for all 186 objects in our sample.  We also compare our 
$H$-band bar classifications with the optical classifications of 
\markcite{sba}Sandage \& Bedke (1994; CAG hereafter) for the 166 objects in 
common with our sample.  We present the statistics of our classifications in 
Table 1.  The two optical catalogs differ with respect to the classification of 
weak bars, representing the extremes of approach to the issue.  The 
\markcite{sba}CAG generally classifies a galaxy as unbarred unless the presence 
of a bar is very clear, whereas the \markcite{rc3}RC3 tends to classify 
galaxies as weakly barred if there is even the slightest evidence for a bar.  

\subsection{Bar Frequency in the $H$-band}

We inspected the $H$-band images of the 186 galaxies in our sample, and 
assigned a Hubble type to each.  We used the standard, subjective 
classification criteria described in the \markcite{sba}CAG and the 
\markcite{rc3}RC3.  We made these classifications with no prior knowledge of 
the optical classifications of the sample (except, of course, for the Survey 
definition).  As this paper is concerned only with bar frequency, we defer a 
full discussion of our classification scheme to \markcite{bap}Eskridge et 
al.~(1999).  We adopt the \markcite{rc3}RC3 distinction between strongly barred 
(SB), weakly barred (SAB), and unbarred (SA) galaxies, although it is clear 
from the results presented below, we were more conservative in our assignment 
of galaxies as weakly barred than is the \markcite{rc3}RC3.  We note that the 
\markcite{rc3}RC3 classification of weakly barred (SAB) differs from the 
\markcite{sba}CAG classification of intermediate barred (S/SB).  As the terms 
``barred,'' ``unbarred,'' ``weakly barred'' and ``intermediate barred'' are not 
explicitly defined in the optical catalogs, we provide examples in Figure 1 of 
$B$-band and $H$-band images of three galaxies: NGC 3223 (unbarred in the 
$H$-band, the \markcite{rc3}RC3, and the \markcite{sba}CAG); NGC 1371 (weakly 
barred in the $H$-band and the \markcite{rc3}RC3, unbarred in the 
\markcite{sba}CAG); NGC 613 (strongly barred in the $H$-band, the 
\markcite{rc3}RC3, and the \markcite{sba}CAG).

One of us (PBE) classified the entire sample twice, with excellent overall 
agreement between the two trials: more than 80\% of the classifications agreed 
to within two subtypes (i.e., Sa to Sb).  We also cross-checked our results by 
comparing classifications done by two of us (PBE \& JAF) for forty galaxies.  
Again, more than 80\% of the tested galaxies were assigned types that agreed to 
within two subtypes.  This compares well with the scatter between classifiers 
reported by \markcite{nea}Naim et al.~(1995).  We note that we do not 
explicitly consider the scale of the bar when making these classifications; 
galaxies with short, high-contrast bars are considered barred.  As the optical
catalogs are based on photographic images, there is some chance of a bias 
against finding short, high-contrast bars in these works.  We discuss this
point in more detail in \S 3.2, below.

\subsection{Comparison with Optical Bar Types}

The bar classification from the \markcite{rc3}RC3 is given for our sample in
Table 1.  Figure 2a shows the distribution of our sample in a 3$\times$3 matrix 
of the \markcite{rc3}RC3 versus our $H$-band bar classification.  We classify 
relatively few galaxies as weakly barred compared to the \markcite{rc3}RC3; 
bars are typically detected in the optical if they are present, but they 
appear stronger in the near-IR. Of the 105 galaxies we classify as strongly 
barred in the $H$-band, there are 10 that are classified as unbarred (SA) in 
the \markcite{rc3}RC3.  We list these galaxies in Table 2.  Most often, these 
are systems with relatively short, high-contrast $H$-band bars that are hidden 
in the optical by patchy foreground extinction and complex inner arm structure. 
In Figure 3 we show $B$- and $H$-band images of two galaxies with optically 
hidden bars, NGC 5161 and NGC 5085.  In \S 4 we shall revisit the topic of
optically hidden bars in the context of galaxy morphology.

There are 15 galaxies, listed in Table 3, that have stronger optical bar-types 
in the \markcite{rc3}RC3 than we assign to the $H$-band images (see Fig.~2a).  
We reinspected our $H$-band images of these galaxies, as well as both the 
\markcite{sba}CAG and our own optical CCD images.  Of these 15 galaxies, three 
are low surface-brightness irregular galaxies (NGC 625, ESO 383-G87, and NGC 
7713) that have vaguely rectangular isophotes.  While barred irregular galaxies 
certainly do exist (e.g., the LMC and NGC 6822), we see no evidence for a bar
in any of these three objects, in either the optical or the near-IR.  Three 
galaxies (NGC 1421, A0908-08, and IC 5052) are edge-on spirals.  It is 
notoriously difficult to detect bars in edge-on galaxies based on imaging data 
alone (see \markcite{kmb}Kuijken \& Merrifield 1995), thus we are not troubled 
by the disagreement for these galaxies.  Three of the 15 galaxies are 
classified as strongly barred in the \markcite{rc3}RC3, but we classify them as 
weakly barred in the $H$-band.  Five of the remaining six are classified as 
weakly barred in the \markcite{rc3}RC3 and as unbarred by us.  The final galaxy 
(NGC 4856) is strongly barred in the \markcite{rc3}RC3 and unbarred in our 
classification.  We do not see evidence for bars in our deep optical CCD images 
for these last six galaxies.  We speculate that these galaxies may have been 
misclassified in the \markcite{rc3}RC3 due to the limitations of the 
photographic images available for classification, and note that none of these 
galaxies are classified as barred in the \markcite{sba}CAG.

We also compare our bar classification to the optical classification from 
\markcite{sba}CAG for the 166 galaxies we have in common (see Table 1).  A 
large number of galaxies classified as optically unbarred in the 
\markcite{sba}CAG galaxies show bars in the near-IR.  Fig.~2b shows the 
distribution of our sample in a 3$\times$3 matrix of \markcite{sba}CAG versus 
$H$-band bar classification.  We note that there is only one object classified 
as intermediate (S/SB) in the \markcite{sba}CAG that we classify as unbarred in 
the $H$-band:  IC 5052, noted above.  There are no galaxies classified as 
strongly barred in the \markcite{sba}CAG that we do not also classify as 
strongly barred in the $H$-band.

One potential bias in the optical classifications is the possibility that bars
are undetected in the optical due to the use of photographic images for
classification in the \markcite{rc3}RC3 and \markcite{sba}CAG.  In many of
these images, the central parts of the galaxy are burned out.  Thus, if there
were a small-scale nuclear bar it might have escaped detection by the optical
classifiers.  We checked this possibility by re-examining our $H$-band images 
and the optical images in the \markcite{sba}CAG for all galaxies that we 
classified as barred or weakly barred in the $H$-band, but that were classified 
as unbarred in either the \markcite{rc3}RC3 or the \markcite{sba}CAG.  There 
are ten galaxies, listed in Table 4, that could suffer from this dynamic range
problem.  Of these, only one (NGC 2280) is classified as unbarred in both the
\markcite{rc3}RC3 and \markcite{sba}CAG.  The other nine are classified as
either SB or SAB in the \markcite{rc3}RC3.  For all ten galaxies, we then
examined our OSU Survey $B$-band images to see if the $H$-band bar was
detectable in high dynamic range optical data.  In five of the ten galaxies no
bar is visible in our deep $B$-band images.  In two cases, the bar is clearly
visible (NGC 4448, NGC 4699).  In two further cases (NGC 3893, NGC 4781) the
bar is visible, but we would not consider the galaxy to be barred without the
prior knowledge of the $H$-band bar.  The remaining galaxy is NGC 1317, which
has both a nuclear bar and a large scale bar.  The nuclear bar is weakly 
visible in our $B$-band image.  Thus we have a few examples of galaxies that
are classified as unbarred in the \markcite{sba}CAG (but not in the 
\markcite{rc3}RC3) that could be misclassified due to the limited dynamic 
range of the optical plates.  We conclude that misclassification of galaxies 
due to the poor dynamic range of the optical classifying plates is not a 
serious bias for our current study.

\section{Bar Fraction as a Function of Morphology}

We have a large enough sample that we are able to examine the bar fraction as
a function of morphology.  In Figure 4 we plot the bar fraction as a function
of morphological type from the \markcite{rc3}RC3 (Fig.~4a), the 
\markcite{sba}CAG (Fig.~4b), and our $H$-band classification (Fig.~4c).  In the
$H$-band, roughly two thirds of the sample show bars, with no significant
dependence on morphology.  In the optical, both the \markcite{rc3}RC3 and the 
\markcite{sba}CAG have a larger bar fraction for the earliest spiral types 
compared to the intermediate types.  This is not statistically significant due
to the small number of galaxies in the earliest-type bin for both the 
\markcite{rc3}RC3 and \markcite{sba}CAG.  However, if one adopts the hypothesis 
that bars are structures in the old stellar population, such an optical excess 
in the earliest types would be expected due to the relative lack of dust and 
young stars in early-type spirals, compared with later types.  The difference 
between the \markcite{rc3}RC3 and \markcite{sba}CAG in the latest types can be 
ascribed entirely to the small number of very late-type systems in our sample 
according to the \markcite{sba}CAG.  The \markcite{rc3}RC3 and 
\markcite{sba}CAG agree that roughly a third of intermediate spirals (Sa 
through Scd) have strong bars.  In the $H$-band, however, about 54\% of 
intermediate spirals have strong bars.

Because the \markcite{rc3}RC3 and the \markcite{sba}CAG treat weak bars so
differently, we examine the influence of morphology on the issue of optically 
hidden bars in two ways.  First, we examine the fraction of unbarred galaxies 
(SA) in the two optical catalogs that we find to have $H$-band bars (SB and 
SAB).  Second, we examine the fraction of galaxies that are not strongly barred 
(SA and SAB) in the two optical catalogs that we find to be strongly barred 
(SB) in the $H$-band.  In Figure 5 we plot the fraction of optically unbarred 
galaxies that have bars in the $H$-band, as a function of morphology, for the 
two optical catalogs.  The solid lines show the fraction of strong $H$-band 
bars, while the dashed lines include weak $H$-band bars.  When we compare with 
the \markcite{rc3}RC3 (Fig.~5a), we find that roughly half of optically 
unbarred spiral galaxies (types Sa -- Scd) reveal the presence of a bar in the 
$H$-band.  The fraction increases toward later types.  As noted earlier, this 
is most likely due to the increasing obscuration due to dust, and the 
increasing prominence of young population tracers in the later-type spirals.  
Comparing, instead, with the \markcite{sba}CAG (Fig.~5b), we find a larger 
fraction of hidden bars, and no evidence for a trend with morphological type.  
About two thirds of the optically unbarred galaxies with types between Sa and 
Scd in the \markcite{sba}CAG have bars in the $H$-band.  In Figure 6 we plot 
the fraction of galaxies with strong $H$-band bars that are not classified as 
strongly barred in the optical.  Here, the solid lines show the results 
including only the optically unbarred galaxies, while the dashed lines include 
the optically weakly barred galaxies.  When we compare with the 
\markcite{rc3}RC3 (Fig.~6a), we find that roughly 40\% of spirals without a 
strong optical bar reveal a strong bar in the $H$-band, with no strong evidence 
for morphological dependence.  Comparing with the \markcite{sba}CAG (Fig.~6b), 
we find essentially the same result.

\section{Summary and Discussion}

We have performed the first analysis of the bar fraction in the near-IR
($H$-band) of a large (186 galaxies) statistically well-defined sample of 
spiral galaxies drawn from the Ohio State Bright Spiral Galaxy Survey.  The 
central result of this work is that the fraction of strongly barred galaxies is 
much higher in the near-IR (nearly 60\% of our sample) than previously found in 
the optical (roughly 30\% of our sample).  In detail, we find 56\% of the 
sample (105 galaxies) to be strongly barred in the $H$-band and 72\% (134 
galaxies) to be barred at some level.  We note that this agrees well with the 
results of \markcite{mnr}Mulchaey \& Regan (1997) and \markcite{ksp}Knapen et 
al.~(1999) for much smaller samples.  For the same sample of objects, the 
\markcite{rc3}RC3 finds 34\% to be strongly barred and 64\% to be either 
strongly or weakly barred.  Thus the major difference between our $H$-band 
classification and the \markcite{rc3}RC3 optical classification is that we find 
a much larger fraction of the sample to be strongly barred.  Only $\sim$20\% of 
the galaxies classified as unbarred (SA) in the \markcite{rc3}RC3 have strong 
(SB) $H$-band bars; this is not a function of morphology.  Roughly 40--50\% of 
galaxies classified as unbarred (SA) in the \markcite{rc3}RC3 have detectable 
bars (SAB \& SB) in the $H$-band.  This fraction increases from $\sim$30\% for 
Sa-Sab galaxies up to $\sim$55\% for Sc-Scd galaxies.  We speculate that this 
morphological effect is the result of the increased influence of dust and young 
stars on the optical morphology of late-type spirals compared to earlier types.

The \markcite{sba}CAG gives classifications for 166 of the galaxies in our
sample.  Of these, it classifies 27\% as strongly barred and 31\% as barred at
any level.  Thus, compared to the \markcite{sba}CAG, we find a much higher
fraction of galaxies with bars.  This is mainly due to the tendency of the
\markcite{sba}CAG to classify galaxies with weak optical bars as unbarred.
Because of this, the fraction of galaxies that are classified as unbarred in 
the \markcite{sba}CAG for which we find $H$-band bars is much larger than for
the \markcite{rc3}RC3:  Nearly 50\% of galaxies that are classified as unbarred
in the \markcite{sba}CAG are strongly barred in the $H$-band.  This number
climbs to $\sim$60\% when we include weak $H$-band bars.  We find no evidence 
for any dependence on morphology in the \markcite{sba}CAG hidden bars.

Although there are galaxies with optically hidden bars, visible only in the
near-IR, we disagree with claims found in the recent literature that all or 
nearly all bright spirals are barred in the near-IR.  The most complete recent 
study that comes to this conclusion is that of \markcite{snj}Seigar \& James 
(1998), who found a bar fraction of 90\% in a non-statistical sample of spiral 
galaxies.  From this they argue that, ``this indicates that most, and quite 
possibly all, bright spirals are barred at some level.''  Clearly, we disagree, 
as we find no evidence for a bar in nearly 30\% of our sample.  It is tempting 
to ascribe the difference between our conclusion and theirs to sample 
differences; our sample is both larger and statistically better-defined.  
However, there are other differences between the two studies that are worth 
consideration.  First, our data are typically 1 magnitude deeper in surface 
brightness than are those of \markcite{snj}Seigar \& James (1998):  Our 
observing times are comparable to theirs; they used the 3.8m UKIRT telescope, 
with a pixel size of $0\rlap.{''}29$; our data were obtained mostly with a 1.8m 
telescope and a detector with $1\rlap.{''}5$ pixels.  As we are concerned with 
extended objects, the surface-brightness limit scales linearly with both the 
aperture and the pixel size.  Thus our images will typically probe a factor of
$\approx$2.5, or about 1 magnitude, deeper in surface brightness.  On this 
basis, we should be more sensitive to the presence of bars than are 
\markcite{snj}Seigar \& James.  Second, \markcite{snj}Seigar \& James consider 
any oval distortion of the central region of a galaxy to be a bar.  Third, they 
attempt full profile decompositions, rather than our ``classical,'' qualitative 
method.  As a result, they may be more sensitive to weak bars.  However, the 
well known failures of profile-fitting algorithms to find disks in some S0s 
suggest some caution on this point (see, e.g., \markcite{mnm}Michard \& Marchal 
1994).  Finally, because they subtract models for the bulge and disk light 
before looking for bars, their method is very sensitive to how well the disk 
and bulge light distributions were modeled.  For these reasons, we do not find 
the differences in the results of the two studies surprising.  We believe that 
our results provide the most representative sample currently available of the 
bar fraction of galaxies in the nearby Universe.

We find that late-type spirals (Sc-Sm) have essentially the same bar fraction 
as early-type spirals (Sa-Sb).  This is consistent with recent results on disk 
shape as a function of Hubble type (\markcite{rnr}Rudnick \& Rix 1998; 
\markcite{znr}Zaritsky \& Rix 1997), arguing that both bars and disk 
asymmetries are due to the dynamical properties of disks.  However, one might 
expect that bar lifetimes in the late-type spirals would be shorter than in 
early-type spirals for at least two reasons.  First, as the disks of later-type 
spirals have a much higher gas fraction than do those of earlier-type spirals, 
angular momentum transfer should operate more strongly in later-type spirals.  
Second, dark halos appear to dominate the dynamics of later-type spirals (e.g.,
\markcite{cnr}Courteau \& Rix 1999) much more than they do 
earlier-type spirals (e.g., \markcite{mgk}Moriondo, Giovanardi \& Hunt 1998).  

Dynamical models of disk galaxies indicate that bars are transient phenomena, 
and can be induced by external interactions (e.g., \markcite{bv}Byrd \& 
Valtonen 1990; \markcite{gca}Gerin, Combes \& Athanassoula 1990).  This 
suggests that the bar fraction of field and cluster spirals could be different.
Although optical studies (e.g., \markcite{vic}Andersen 1996; 
\markcite{eeb}Elmegreen, Elmegreen \& Bellin 1990) find no compelling evidence 
for any such difference, it is worth examining our sample to see if any 
difference arises in the near-IR.  Given the magnitude limit of our survey, the 
only clusters represented are Fornax and Virgo.  There are a total of 19 
galaxies in our sample that are members of these clusters according to 
\markcite{frx}Ferguson (1989) and \markcite{vcc}Binggeli, Sandage \& Tammann 
(1985).  Of these, we classify 12 (63\%) as barred, three (16\%) as weakly 
barred, and four (21\%) as unbarred.  Thus we find that the fraction of barred 
galaxies in the Fornax and Virgo clusters is slightly higher than for our full 
sample, but the total numbers are small enough that this is not statistically 
significant.  

A number of recent studies have attempted to evaluate the evolution of the
distribution of Hubble types with increasing redshift (e.g., 
\markcite{lea}Lilly et al.~1998; \markcite{dea}Driver et al.~1998; 
\markcite{sea}Simard et al.~1999), but these studies generally avoid the issue 
of bars.  \markcite{aea}Abraham et al.~(1999) have recently claimed that the 
fraction of barred galaxies decreases with increasing redshift, becoming nearly 
zero beyond a redshift of $z \approx 0.5$.  This apparent lack of bars at high 
redshift may be a crucial piece of information for understanding galaxy 
evolution.  However, this result may be driven by a combination of systematics,
involving both the band-pass shift of the rest frame with increasing redshift
and the younger stellar populations in galaxy disks at increasing redshifts.  
In keeping with this, Bunker (1999) points out that several optically unbarred 
galaxies in the Hubble Deep Field (HDF) reveal bars in the NICMOS HDF data.  
Although \markcite{aea}Abraham et al.~claim that the bandpass effect is 
insufficient to account for their result, it may be that the combination of the 
shift to bluer rest wavelengths, and the younger mean stellar population of 
galaxy disks at look-back times of 4--7 Gyr is sufficient to account for the 
lower detection frequency of bars in the \markcite{aea}Abraham et al.~sample.  
A careful modeling study of these effects should shed considerable light on 
this issue.

In our next paper we shall present a full discussion of the optical and near-IR
morphologies of the galaxies in our sample, based on our $B$- and $H$-band 
images.  This will allow us to consider the strength of features such as rings 
as a function of wavelength.  Returning to the specific issue of bars, we note
that ``bar strength'' does not map onto a single physical variable (at least
two are involved:  relative flux at a given wavelength, and bar axial ratio).  
A number of different quantitative measures of ``bar strength'' have been 
proposed in the literature (e.g., \markcite{e2}Elmegreen \& Elmegreen 1985; 
\markcite{ohw}Ohta, Hamabe \& Wakamatsu 1990; \markcite{m95}Martin 1995; 
\markcite{woz}Wozniak et al.~1995; \markcite{aea}Abraham et al.~1999), all of 
which measure different things.  We will apply a collection of these various 
measures to our sample in an attempt to put the phrase ``bar strength'' on a 
firm, consistent, quantitative footing.  As part of this study, we also plan to 
examine the isophotal shapes of bulges in barred and unbarred spirals.  

\acknowledgments

We thank the many OSU graduate students who collected data for this project, 
and wish to especially note the many nights of work that Ray Bertram and Mark
Wagner have devoted to the OSU survey.  We are grateful to Roberto Aviles for 
obtaining many of the optical images of southern galaxies for us with the 0.9 
meter telescope at CTIO.  We thank Bob Williams and Malcolm Smith, past and 
present directors of CTIO, for the generous allotment of telescope time needed 
to observe most of our southern sample.  JAF thanks Leonard Searle for the 
observing opportunites provided by a Visiting Research Associateship at Las 
Campanas Observatory, where some of these data were obtained.  We are pleased 
to thank Ron Buta and Allan Sandage for several very useful discussions on 
galaxy morphology, and Roelof deJong for his comments.  JAF acknowledges 
support from a PPARC Senior Visiting Research Fellowship (grant no.~GR/L00896) 
held in 1996 at the University of Durham Physics Department.  This work was 
supported by grants AST-9217716 and AST-9617006 from the National Science 
Foundation.

\newpage

{
\baselineskip12pt
\tolerance=500
 
\def\tabrule{\noalign{\hrule}}
\def\pz{\phantom{0}}
\def\pb{\phantom{-}}
\def\pd{\phantom{.}}
\ 
 
\centerline{Table 1 - Bar Fraction in the $H$-band and Optical Catalogs}
\vskip0.3cm
 
\newbox\tablebox
\setbox\tablebox = \vbox {
 
\halign{\pz\pz#\pz\pz&\hfil\pz\pz#\pz\pz\hfil&\hfil\pz\pz#\pz\pz\hfil&\hfil\pz
\pz#\pz\pz\hfil&\hfil\pz\pz#\pz\pz\hfil&\hfil\pz\pz#\pz\pz\hfil&\hfil\pz\pz#\pz
\pz\hfil\cr
\tabrule
\noalign{\vskip0.1cm}
\tabrule
\noalign{\vskip0.1cm}

Bar Class & $H$-band & & RC3 & & CAG & \cr
\ & Fraction & Number & Fraction & Number & Fraction & Number \cr 
\noalign{\vskip0.1cm}
\tabrule
\noalign{\vskip0.2cm}
SB & 56\% & 105 & 35\% & $\pz$65 & 27\% & $\pz$44 \cr
SAB & 16\% & $\pz$30 & 30\% & $\pz$56 & $\pz$4\% & $\pz\pz$6 \cr
SB + SAB & 73\% & 135 & 65\% & 121 & 30\% & $\pz$50 \cr
SA & 27\% & $\pz$51 & 35\% & $\pz$65 & 70\% & 116 \cr
SA + SAB & 44\% & $\pz$81 & 66\% & 121 & 74\% & 122 \cr
\noalign{\vskip0.2cm}
\tabrule
}
}
\centerline{ \box\tablebox}
\noindent{In columns 2, 4, and 6 we give the fraction of galaxies in our sample 
with a given bar class in the $H$-band, the \markcite{rc3}RC3, and the 
\markcite{sba}CGA, respectively.  Columns 3, 5, and 7 give the total number of 
galaxies in our sample in each bar class from the same sources.}
}

\vskip20pt
 
{
\baselineskip12pt
\tolerance=500
 
\def\tabrule{\noalign{\hrule}}
\def\pz{\phantom{0}}
\def\pb{\phantom{-}}
\def\pd{\phantom{.}}
\ 
 
\centerline{Table 2 - Strongly IR Barred Galaxies, Unbarred in RC3}
\vskip0.3cm
 
\newbox\tablebox
\setbox\tablebox = \vbox {
 
\halign{\hfil\pz\pz#\pz\pz\hfil&\hfil\pz\pz#\pz\pz\hfil&\pz\pz#\pz\pz\hfil\cr
\tabrule
\noalign{\vskip0.1cm}
\tabrule
\noalign{\vskip0.1cm}

 & $H$-band type & RC3 type \cr
\noalign{\vskip0.1cm}
\tabrule
\noalign{\vskip0.3cm}
NGC 1511 & SBd?$\pz$ & SAa: pec \cr
NGC 2280 & SBbc$\pz$ & SA(s)cd \cr
NGC 3675 & SB(r)a & SA(s)b \cr
NGC 4388 & SB(r)a & SA(s)b: sp \cr
NGC 4450 & SBab$\pz$ & SA(s)ab \cr
NGC 4504 & SBcd$\pz$ & SA(s)cd \cr
NGC 5161 & SBab$\pz$ & SA(s)c: \cr
NGC 5483 & SBc$\pz\pz\pd$ & SA(s)c \cr
NGC 5962 & SBab$\pz$ & SA(r)c \cr
NGC 6902 & SB(r)a & SA(r)b \cr
\noalign{\vskip0.2cm}
\tabrule
}
}
\centerline{ \box\tablebox}
}
 
\vskip20pt

{
\baselineskip12pt
\tolerance=500
 
\def\tabrule{\noalign{\hrule}}
\def\pz{\phantom{0}}
\def\pb{\phantom{-}}
\def\pd{\phantom{.}}
\ 
 
\centerline{Table 3 - Galaxies with Stronger RC3 Bars than $H$-band Bars}
\vskip0.2cm
 
\newbox\tablebox
\setbox\tablebox = \vbox {
 
\halign{\hfil\pz\pz#\pz\pz\hfil&\hfil\pz\pz#\pz\pz\hfil&\hfil\pz\pz#\pz\pz\hfil
\cr
\tabrule
\noalign{\vskip0.1cm}
\tabrule
\noalign{\vskip0.1cm}

$H$-band SA, RC3 SAB & $H$-band SA, RC3 SB & $H$-band SAB, RC3 SB \cr
\noalign{\vskip0.1cm}
\tabrule
\noalign{\vskip0.2cm}
NGC $\pz$157 & NGC $\pz$625$\pz\pz\pz\pd$ & NGC 1617 \cr
NGC $\pz$278 & NGC 4856$\pz\pz\pz\pd$ & NGC 1703 \cr
NGC 1421 & ESO 383$-$G87 & NGC 7412 \cr
NGC 1964 & IC$\pb$ $\pz$5052$\pz\pz\pz$ & \cr
A0908$-$08 & NGC 7713$\pz\pz\pz\pd$ & \cr
NGC 4580 & & \cr
NGC 5248 & & \cr 
\noalign{\vskip0.2cm}
\tabrule
}
}
\centerline{ \box\tablebox}
}

\vskip20pt
 
{
\baselineskip12pt
\tolerance=500
 
\def\tabrule{\noalign{\hrule}}
\def\pz{\phantom{0}}
\def\pb{\phantom{-}}
\def\pd{\phantom{.}}
\ 
 
\centerline{Table 4 - IR Barred Galaxies with Potential Optical Dynamic Range 
Problems}
\vskip0.3cm
 
\newbox\tablebox
\setbox\tablebox = \vbox {
 
\halign{\hfil\pz\pz#\pz\pz\hfil&\hfil\pz\pz#\pz\pz\hfil\cr
\tabrule
\noalign{\vskip0.1cm}
\tabrule
\noalign{\vskip0.1cm}

Unbarred in $B$-band CCD & Barred in $B$-band CCD \cr
\noalign{\vskip0.1cm}
\tabrule
\noalign{\vskip0.2cm}
NGC $\pz$779 & NGC 1317 \cr
NGC 2280 & NGC 3893 \cr
NGC 3166 & NGC 4448 \cr
IC $\pz\pz$4444 & NGC 4699 \cr
IC $\pz\pz$5325 & NGC 4781 \cr
\noalign{\vskip0.2cm}
\tabrule
}
}
\centerline{ \box\tablebox}
}
 
\newpage

\figcaption{$B$- and $H$-band images of a) NGC 3223 (unbarred in the $H$-band, 
the RC3, and the CAG), b) NGC 1371 (weakly barred in the $H$-band and the RC3, 
unbarred in the CAG), and c) NGC 613 (strongly barred in the $H$-band, the RC3, 
and the CAG).}

\figcaption{Grids showing the distribution of galaxies in cells according to 
a) the RC3 and our $H$-band bar classification, b) the CAG and our $H$-band bar 
classification.}

\figcaption{$B$- and $H$-band images of a) NGC 5161, and b) NGC 5085.  These
are galaxies classified as optically unbarred (SA) in the RC3, but as 
barred (SB and SAB, respectively) in the $H$-band by us.}

\figcaption{Bar-fraction as a function of morphological type: a) optical types 
and bar classification from the RC3; b) optical types and bar classification 
from the CAG; c) $H$-band types and bar classification.  The solid lines show
the fraction of strong bars (SB).  The dashed lines show the results of 
including weak bars (SAB or S/SB).  The numbers above the histogram bins are
the number of barred galaxies in that bin (including the weakly barred
galaxies in parentheses) out of the total number of galaxies in that bin.}

\figcaption{Fraction of optically unbarred galaxies with infrared bars plotted
as a function of morphology:  a) optical types and bar classification from the 
RC3; b) optical types and bar classification from the CAG.  The solid lines 
show the fraction of strong bars (SB).  The dashed lines show the results 
including weak bars (SAB or S/SB).  The numbers above the histogram bins have 
the same meaning as in Figure 4.}

\figcaption{Fraction of galaxies not classified as strongly barred in the
optical, but that are strongly barred in the $H$-band, plotted as a function of 
morphology:  a) optical types and bar classification from the RC3; b) optical 
types and bar classification from the CAG.  The solid lines show the fraction 
of galaxies that are optically unbarred (SA or S).  The dashed lines show the 
results including weak bars (SAB or S/SB).  The numbers above the histogram 
bins have the same meaning as in Figure 4.}

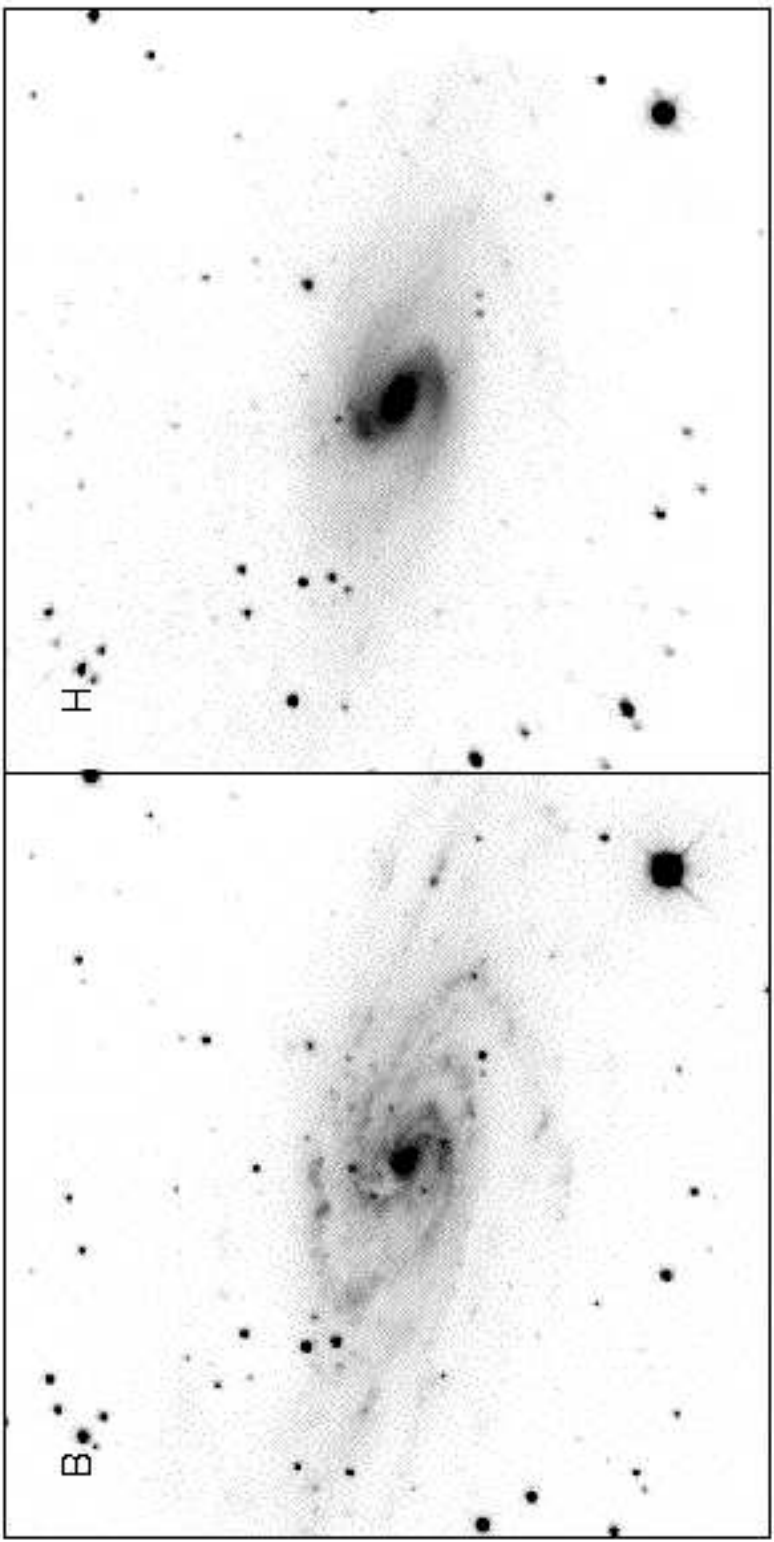
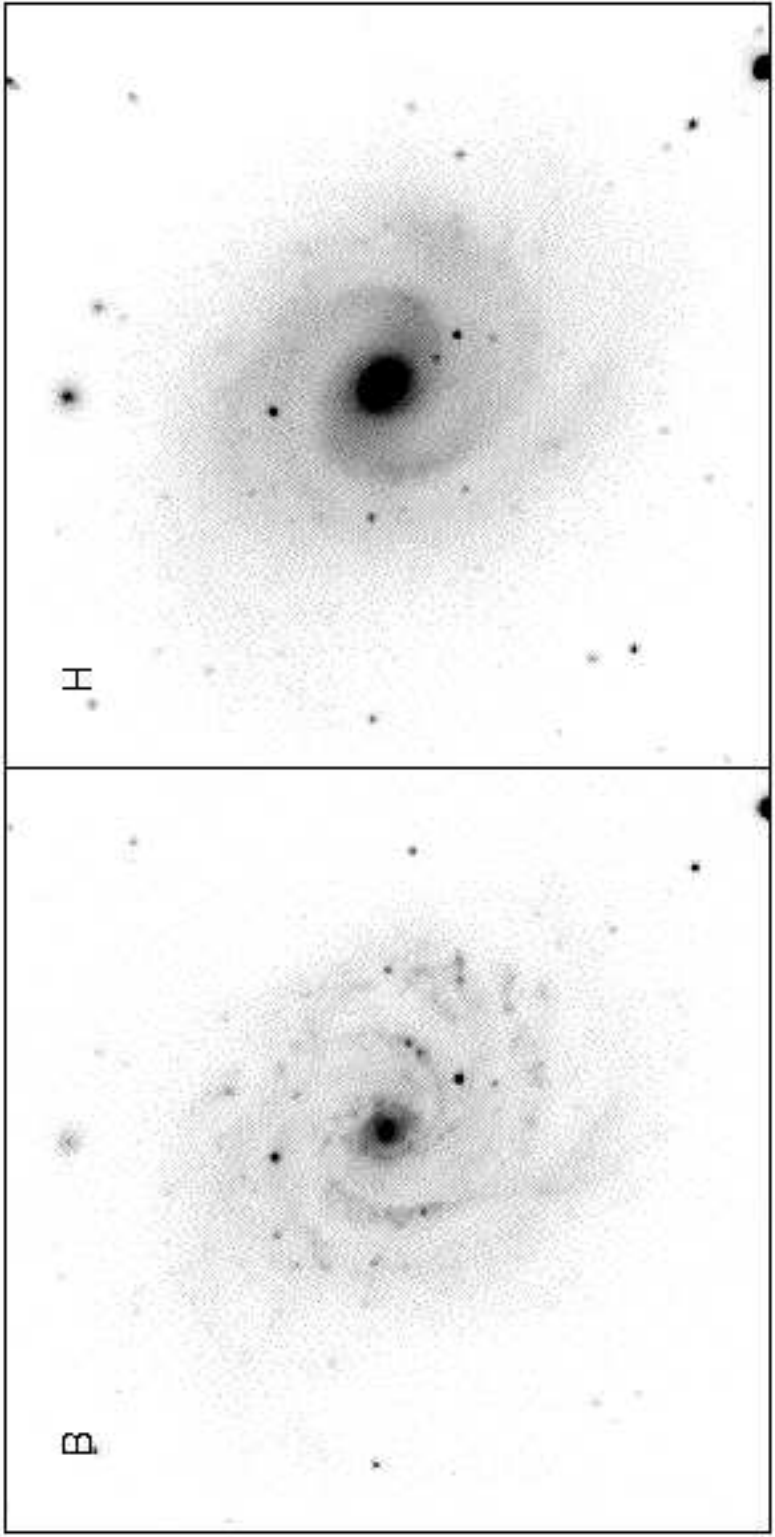
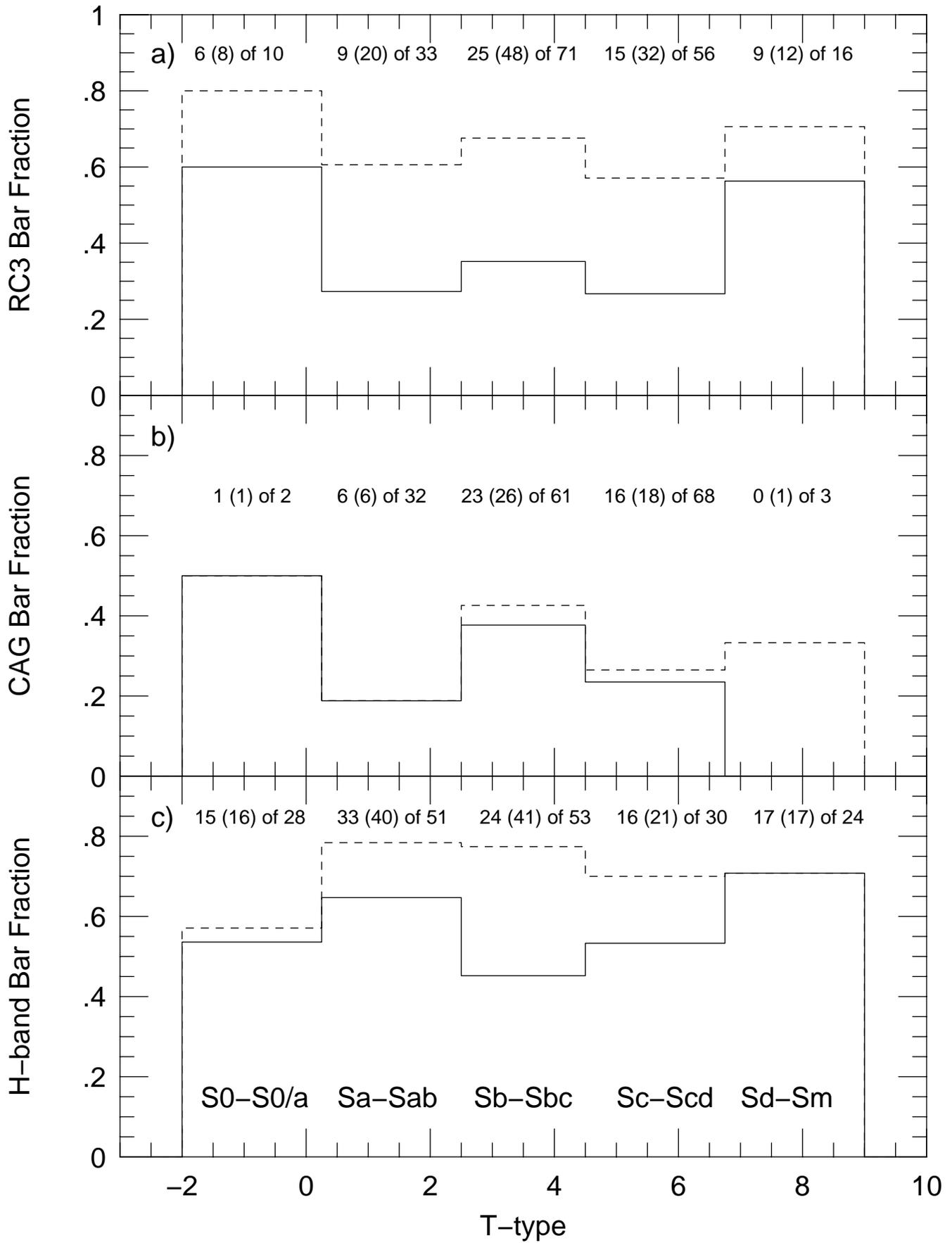
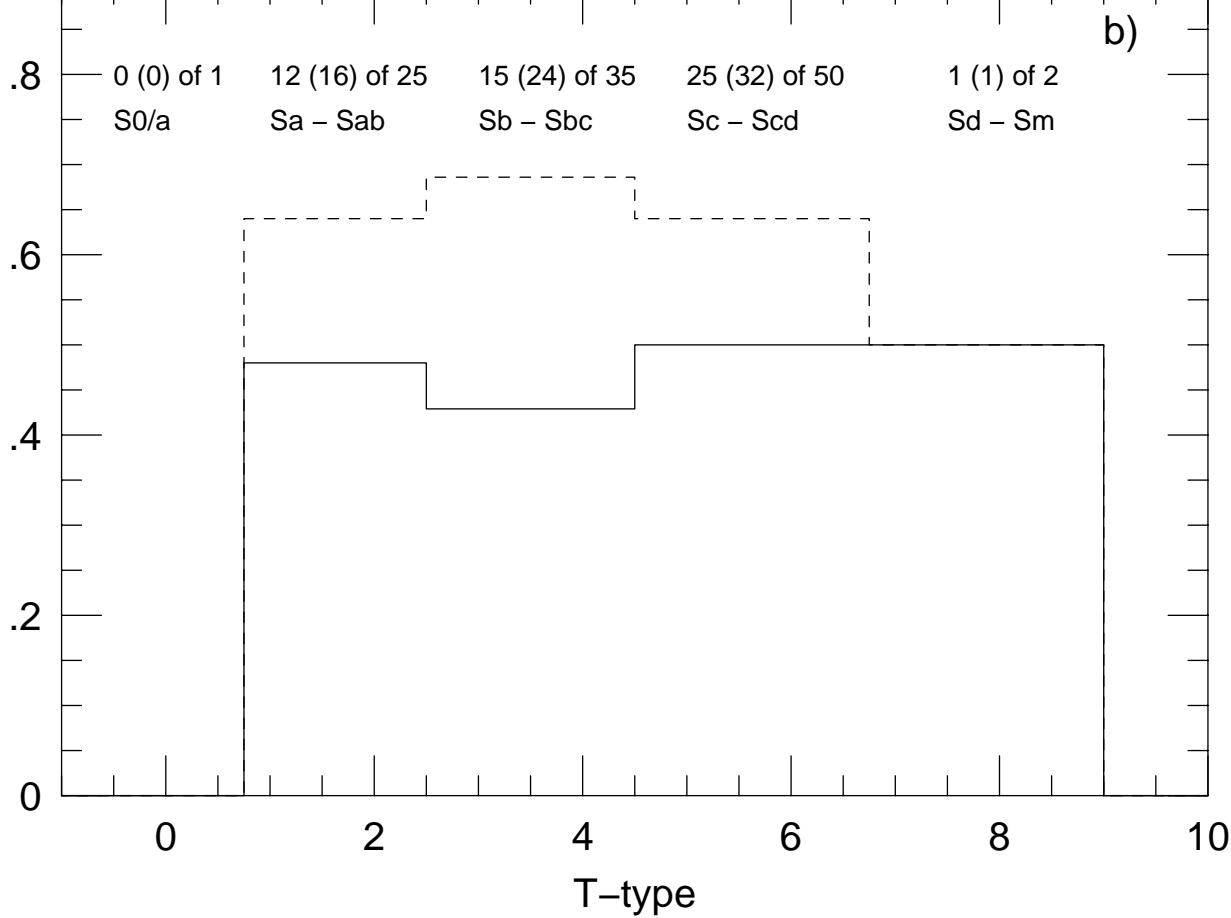
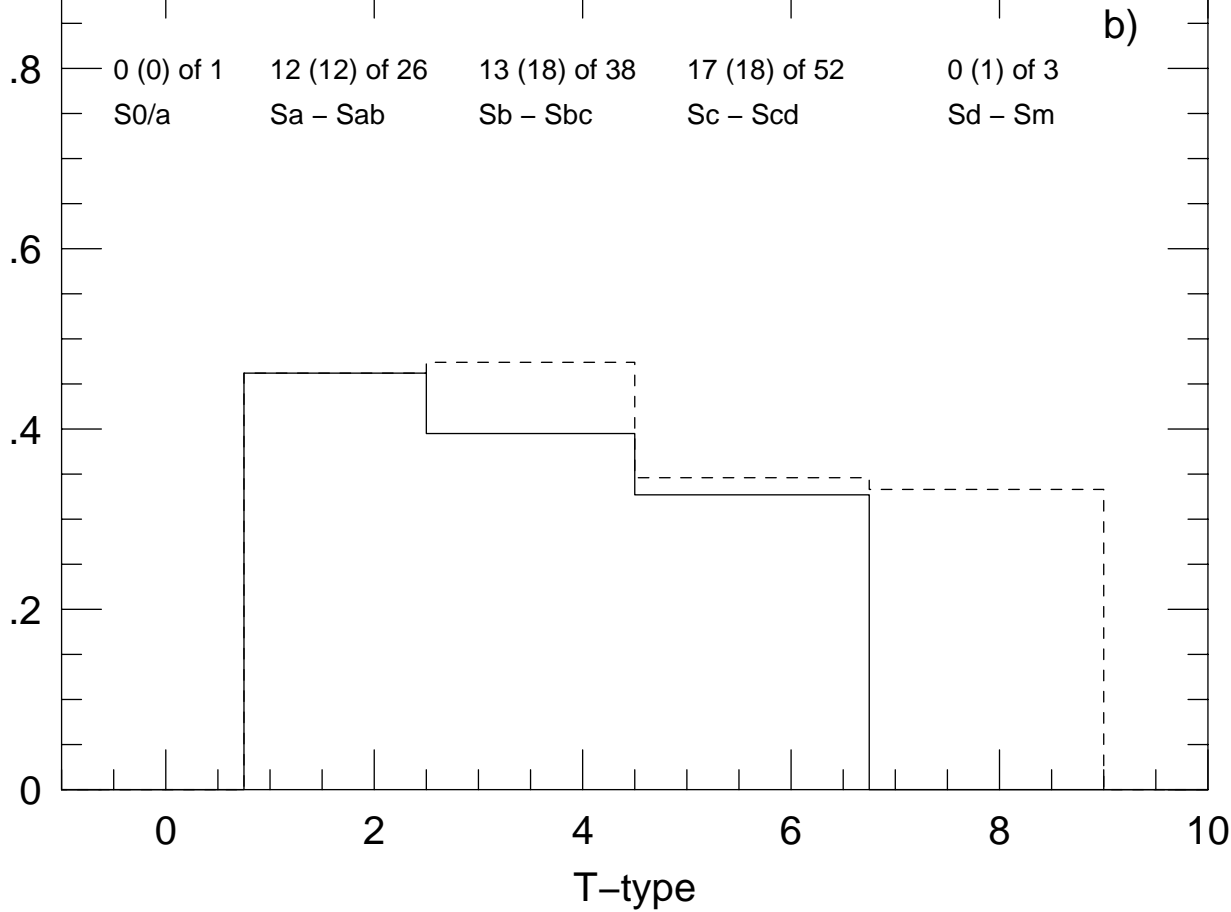
\end{document}